\documentclass[preprint]{elsarticle}

\usepackage{lineno}
\usepackage{hyperref}
\modulolinenumbers[5]

\usepackage[british]{babel}

\journal{Chemical Physics}

\hypersetup{colorlinks=true,linkcolor=blue,citecolor=blue,urlcolor=blue}

\usepackage{amsmath}
\usepackage{bm}
\usepackage{braket} %
\newcommand{\ketbra}[2]{\ket{#1}\!\bra{#2}}

\newcommand{\eu}{\mathrm{e}^}
\newcommand{\iu}{\ensuremath{\mathrm{i}}}
\newcommand{\rmd}{\mathrm{d}}
\newcommand{\thalf}{{\ensuremath{\tfrac{1}{2}}}} %
\newcommand{\half}{{\ensuremath{\frac{1}{2}}}}

\newcommand{\op}[1]{\ensuremath{\hat{#1}}}
\providecommand{\mat}[1]{\mathsf{#1}}
\renewcommand{\mathbf}[1]{\bm{#1}}

\DeclareMathOperator{\sgn}{sgn}
\DeclareMathOperator{\Tr}{Tr}
\DeclareMathOperator{\tr}{tr}

\newcommand{\T}{{\mathstrut\top}}

\newcommand{\pder}[3][]{\frac{\partial^{#1}{#2}}{\partial{#3}^{#1}}}

\newcommand{\Eqn}[1]{Equation\,(\ref{#1})}
\newcommand{\eqn}[1]{Eq.\,(\ref{#1})}
\newcommand{\eqs}[1]{Eqs.\,(\ref{#1})}
\newcommand{\fig}[1]{Fig.\,\ref{fig:#1}}
\newcommand{\secref}[1]{Sec.\,\ref{sec:#1}}
\newcommand{\Appendix}{\ref{sec:appendix}}
\newcommand{\Ref}[1]{Ref.~\cite{#1}}
\newcommand{\Ne}{\Lambda}

\begin{document}

\begin{frontmatter}

\title{An analysis of nonadiabatic ring-polymer molecular dynamics and its application to vibronic spectra}

\author[erlangen,durham]{Jeremy O. Richardson\fnref{eth}}
\ead{jeremy.richardson@fau.de}

\fntext[eth]{Present address: Laboratorium f\"ur Physikalische Chemie, ETH Z\"urich, Switzerland}

\author[erlangen]{Philipp Meyer\fnref{fn1}}

\author[erlangen]{Marc-Oliver Pleinert\fnref{fn1}}

\author[erlangen]{Michael Thoss}

\fntext[fn1]{These authors contributed equally}

\address[erlangen]{Institut f{\"u}r Theoretische Physik und Interdisziplin{\"a}res Zentrum f{\"u}r Molekulare Materialien,
Friedrich-Alexander-Universit{\"a}t Erlangen-N{\"u}rnberg (FAU),
Staudtstrasse 7/B2,
91058 Erlangen, Germany}

\address[durham]{Department of Chemistry, University of Durham, South Road, Durham, DH1 3LE, UK}

\begin{abstract}
Nonadiabatic ring-polymer molecular dynamics
employs the mapping approach to describe nonadiabatic effects within the ring-polymer ansatz.
In this paper, it
is generalized to allow
for the nuclear and electronic degrees of freedom to be described by different numbers of ring-polymer beads.
Analysis of the resulting method 
shows that as the number of electronic mapping variables increases,
certain problems associated with the approach are removed,
such as the non-unique choice of the mapping Hamiltonian
and negative populations leading to inverted potential-energy surfaces.
Explicit integration over cyclic variables
reduces the sign problem for the initial distribution in the general case.
A new application for the simulation of vibronic spectra
is described and promising results are presented for a model system.
\end{abstract}

\begin{keyword}
vibronic spectra\sep
ring-polymer molecular dynamics\sep
nonadiabatic\sep
mapping approach
\end{keyword}

\end{frontmatter}

\includegraphics[width=\textwidth]{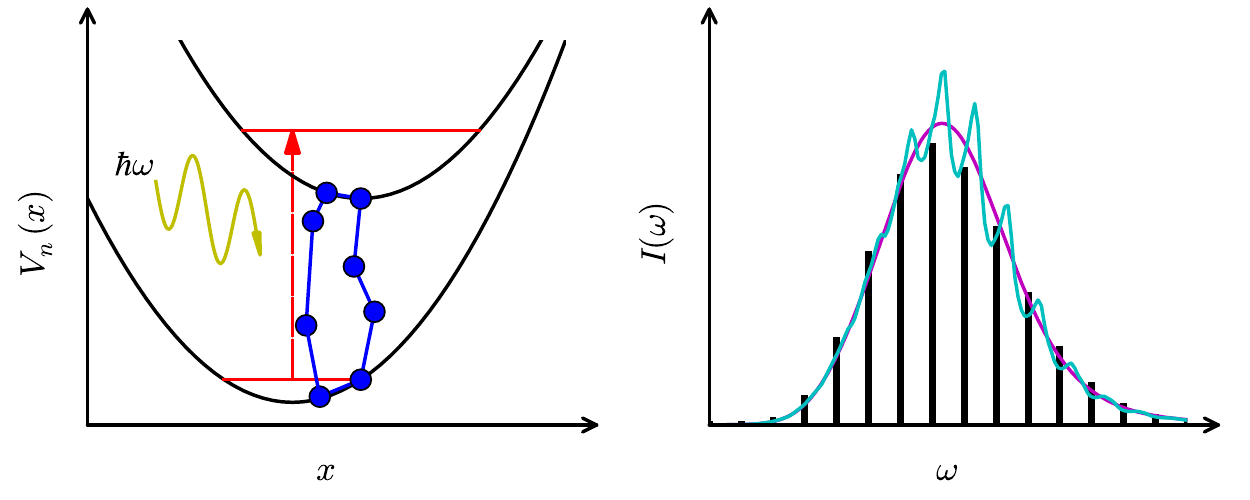}

\section{Introduction}

Nonadiabatic processes, such as those induced by interactions with light or which describe electron
transfer, are present in many areas of physical and biological science including chemical reactions, solar
cells, vision, photosynthesis and DNA radiation damage. 
Modern experimental techniques, including femtosecond
pump-probe spectroscopy as well absorption, emission and photodissociation spectroscopy, have revealed a wealth of information on excited electronic states and their dynamics.
In order to
fully interpret these experimental data, one needs to be able to simulate nonadiabatic dynamics accurately
and efficiently \cite{Tully2012perspective,ConicalIntersections1}.

A number of schemes exist for approximating nonadiabatic dynamics 
such as the Ehrenfest mean-field approach, surface hopping \cite{Tully1998MQC,Shushkov2012RPSH},
classical and semiclassical mapping approaches
\cite{Meyer1979nonadiabatic,Stock1997mapping,Thoss1999mapping,Mueller1998mapping,Stock1999ZPE,Sun1997mapping,Sun1998mapping,Wang1999mapping,Bonella2001mapping2,Thoss2004review,Miller2009mapping},
multiple spawning \cite{Levine2007AIMS},
quantum-classical Liouville dynamics \cite{Kapral2006QCL,Kapral2015QCL},
and other mixed quantum-classical approaches  \cite{Agostini2013MQC}.
Each of these methods has its own advantages and disadvantages
\cite{Stock2005nonadiabatic}
but there still remain problems for which these methods lack either accuracy or efficiency.
The methods presented in this paper
are derived with the intention of balancing accuracy and efficiency
such that nuclear quantum effects can be included
in atomistic simulations of nonadiabatic dynamics in condensed-phase molecular systems.

The method of
ring-polymer molecular dynamics (RPMD)
\cite{RPMDcorrelation,Habershon2013RPMDreview}
has become the method of choice
for including quantum effects into 
many simulations of complex molecular systems.
In particular, its use for computing reaction rates is well established
\cite{rpinst,Hele2013QTST,Althorpe2013QTST,Suleimanov2016rate}.
Early attempts to simulate vibrational spectra using RPMD
discovered artificial resonances contaminating the results
\cite{Habershon2008spectra,Witt2009spectroscopy}
but recent work has shown how to avert this problem
\cite{Rossi2014resonance,Rossi2014spectra}.

The standard RPMD approach is only applicable
within the single-surface Born-Oppenheimer approximation,
and in order to study electron-transfer reactions,
the transferred electron must be explicitly treated as a particle
\cite{Menzeleev2011ET,Kretchmer2013ET}.
One would instead like to treat nonadiabatic problems using a Hamiltonian
with more than one potential-energy surface corresponding to the different electronic states.
Extensions of RPMD to such multi-state Hamiltonians have been suggested
using the mapping approach
\cite{Meyer1979nonadiabatic,Stock1997mapping,Miller2009mapping,Ananth2010mapping,Hele2016Faraday}.
Two such formulations are called nonadiabatic RPMD (NRPMD) \cite{mapping}
and mapping-variable RPMD (MVRPMD) \cite{Ananth2013MVRPMD,Duke2015MVRPMD,Duke2016Faraday}.

Both approaches are defined such that trajectories are initialized
from an exact path-integral quantum distribution and then are evolved using Hamilton's equations of motion.
The dynamics employed in our approach, NRPMD, which we discuss in this paper,
is defined such that Rabi oscillations \cite{MukamelBook} are described exactly for a system
in the limiting case where the electronic states are not coupled to the nuclear positions.
The MVRPMD dynamics are not exact in this limiting case but do in general
conserve their probability distribution, which is not true of NRPMD.

Although it is clear that NRPMD does not conserve its probability distribution
for a finite number of ring-polymer beads,
numerical tests show that the error is reduced as the number of beads increases \cite{mapping}.
It is not yet known what occurs in the infinite limit.
For many cases, conservation of the Boltzmann distribution in this limit would be sufficient
and hence this is an important question that remains currently unanswered.

The goals of this paper are to analyse the NRPMD approach in detail.
In \secref{NRPMD},
we generalize the NRPMD method to allow us to choose more than one set of mapping variables
per ring-polymer bead
and thus control the convergence of the nuclear and electronic degrees of freedom independently.
We analyse the resulting method to understand more about its infinite limit in \secref{analysis}
and show how it can be made more efficient
by partially alleviating the sign problem associated with the initial distribution.
A peculiarity of the mapping approach is that 
there is a non-unique choice which is made in obtaining the mapping Hamiltonian
which can affect the resulting trajectory dynamics.
We show how NRPMD successfully resolves this issue and can be considered unique in this respect.
A new application of the method is described in \secref{vibronic} for computing vibronic spectra,
which compares well with exact results for a model system.

\section{Nonadiabatic RPMD}
\label{sec:NRPMD}

The NRPMD method was first presented in \Ref{mapping}.
Here we follow a derivation similar to that of the previous case
but we include an extension to treat more than one set of mapping variables per ring-polymer bead.
For simplicity, we specify a system with two electronic states
but the method is general and can also be applied to systems with more states.

Nonadiabatic systems have more than one potential-energy surface,
and in the diabatic representation \cite{ConicalIntersections1}
we define a potential-energy matrix with the form
\begin{align}
	\mat{V}(\mat{x}) = \begin{pmatrix} V_0(\mat{x}) & \Delta(\mat{x}) \\ \Delta(\mat{x}) & V_1(\mat{x}) \end{pmatrix} ,
\end{align}
where
$V_n(\mat{x})$ is the diabatic potential-energy surface for the $n$th state
and
$\Delta(\mat{x})$ is the coupling between the states.
There may also be a state-independent potential, $U(\mat{x})$, which affects all electronic states equally.
These functions are scalar fields associated with each point, $\mat{x}$, in multidimensional space representing all nuclear degrees of freedom of the system.
The matrix is in the state space of electronic states, $\ket{n}$ with $n\in\{0,1\}$,
which is complete in the sense that $\sum_{n=0}^1 \ketbra{n}{n}=1$,
such that for instance, $\braket{n|\mat{V}(\mat{x})|n}=V_n(\mat{x})$.

The total Hamiltonian is
\begin{align}
	\label{Hdiabatic}
	\op{H} &= \frac{|\op{\mat{p}}|^2}{2m} + U(\op{\mat{x}}) + \sum_{n=0}^1 V_n(\op{\mat{x}}) \ketbra{n}{n} + \Delta(\op{\mat{x}}) \big(\ketbra{0}{1} + \ketbra{1}{0}\big) \, ,
\end{align}
where $\mat{p}$ are the nuclear momenta and all degrees of freedom have been mass-weighted to have the same mass, $m$.
The partition of the potential into state-dependent and independent parts
is not a unique choice, and
although this would make no difference for the exact quantum solution,
it may affect the approximate trajectory dynamics introduced in \secref{dynamics}.
This is a disadvantage of the classical implementation of the mapping approach which,
as we show in \secref{analysis},
is avoided by NRPMD.

\subsection{Mapping Approach}
\label{sec:mapping}

One reason why efficient nonadiabatic dynamics approaches are difficult to formulate
is because the Hamiltonian includes both continuous and discrete degrees of freedom.
Even if the nuclear dynamics can be described adequately using classical trajectories,
the coupling to discrete electronic states provides an obstacle to a purely classical description \cite{Miller2009mapping,Stock2005nonadiabatic}.

The mapping approach
replaces each electronic state by a fictitious harmonic oscillator degree of freedom.
There is a formal mapping between an electronic state $\ket{n}$
and particular vibrational states of a set of harmonic oscillators.
In this case, the set has two oscillators, 
of which the $n$th is in its first excited state and the
other in its ground state \cite{Thoss1999mapping}.

The wave functions of these states are known
in the position, $\mat{X}=(X_0,X_1)$, and momentum, $\mat{P}=(P_0,P_1)$ bases:
\begin{align}
	\label{wavefunctions}
	\braket{\mat{X}|n} &= \sqrt\frac{2}{\pi} X_n \, \eu{-\half|\mat{X}|^2}
	&
	\braket{\mat{P}|n} &= -\iu \sqrt\frac{2}{\pi} P_n \, \eu{-\half|\mat{P}|^2}
\end{align}
and the correspondence between the operators is
\begin{equation}
	\ketbra{n}{m} \mapsto \op{a}_n^\dag \op{a}_m \, ,
\end{equation}
where $\op{a}_n^\dag=\frac{1}{\sqrt{2}}(\op{X}_n - \iu\op{P}_n)$ and $\op{a}_n=\frac{1}{\sqrt{2}}(\op{X}_n + \iu\op{P}_n)$
are creation and annihilation operators for the $n$th harmonic oscillator.
The electronic state operators can thus be mapped
to functions of the dimensionless position and momentum operators in the fictitious coordinates
as
\begin{subequations}
\label{mapping}
\begin{align}
	\ketbra{n}{n} &\mapsto \thalf (\op{X}_n^2 + \op{P}_n^2 - 1) \label{nn}
\\
	\ketbra{n}{m} + \ketbra{m}{n} &\mapsto \op{X}_n \op{X}_m + \op{P}_n \op{P}_m
	& n &\ne m
	\\
	\iu(\ketbra{n}{m} - \ketbra{m}{n}) &\mapsto \op{P}_n \op{X}_m - \op{X}_n \op{P}_m
	& n &\ne m \, .
\end{align}
\end{subequations}
The final term of \eqn{nn} occurs because $\op{X}_n$ and $\op{P}_n$ do not commute
and $[\op{X}_n,\op{P}_n]=\iu$.
In this way, all degrees of freedom in the system can be written in terms of continuous variables
and the nuclear and electronic states can be treated on an equivalent footing when making classical-like approximations to the dynamics.

Employing the relations in \eqn{mapping},
the Hamiltonian in the mapping representation is
\begin{align}
	\op{\mathcal{H}} = \frac{|\op{\mat{p}}|^2}{2m} + U(\op{\mat{x}}) + \sum_{n=0}^1 \thalf(\op{X}_n^2 + \op{P}_n^2 - 1) V_n(\op{\mat{x}}) + \Delta(\op{\mat{x}}) (\op{X}_0 \op{X}_1 + \op{P}_0 \op{P}_1) \, ,
\end{align}
or equivalently in an alternative notation
\begin{align}
	\label{Hmapping}
	\op{\mathcal{H}} = \frac{|\op{\mat{p}}|^2}{2m} + U(\op{\mat{x}}) + \thalf
		\left[ \op{\mat{X}}^\T\mat{V}(\op{\mat{x}})\op{\mat{X}} + \op{\mat{P}}^\T\mat{V}(\op{\mat{x}})\op{\mat{P}} - \tr\mat{V}(\op{\mat{x}}) \right].
\end{align}

An exact quantum-mechanical solution of the dynamics resulting from the Hamiltonian in \eqn{Hmapping}
would give
the same results as that of \eqn{Hdiabatic}
when interpreted using the correspondences in \eqn{mapping}
\cite{Stock1997mapping}.
The mapping Hamiltonian is however a more useful point from which to take approximations for the dynamics
and trajectory-based methods can be more easily applied.

\subsection{Nonadiabatic ring-polymer statistics}
\label{sec:statistics}

Before we discuss dynamics,
we first derive an approach for obtaining exact quantum statistics using the mapping variables
within a path-integral framework.
The partition function $Z=\Tr[\eu{-\beta\op{H}}]$ can be expanded as a Trotter product as
\begin{align}
	Z
	&\simeq \Tr\left[ \prod_{i=1}^N \eu{-\beta_N |\op{\mat{p}}|^2/2m} \eu{-\beta_N U(\mat{x})} \eu{-\beta_N \mat{V}(\op{\mat{x}})} \right]
	\\
	&= \Tr\left[ \prod_{i=1}^N \eu{-\beta_N |\op{\mat{p}}|^2/2m} \eu{-\beta_N U(\mat{x})} \prod_{\alpha=1}^\Ne \eu{-\beta_N \mat{V}(\op{\mat{x}})/2\Ne} \mathcal{P}^2 \eu{-\beta_N \mat{V}(\op{\mat{x}})/2\Ne} \mathcal{P}^2 \right],
\end{align}
where
$1/k_\mathrm{B}T=\beta=N\beta_N$
and $\mathcal{P} = \sum_{n=0}^1 \ketbra{n}{n}=1$
is a projection operator which
we can introduce an arbitrary number of times without affecting the exact result.
It is used to ensure that the mapping variables are projected onto the correct subspace of singly-excited set of oscillators \cite{Ananth2010mapping}.

Note that here we have split the final factor into $2\Ne$ parts.
This is a slight extension to the method introduced in \Ref{mapping}
which will allow for $\Ne$ sets of mapping variables per nuclear bead.
The original formulation is recovered by setting $\Ne=1$.
The expression for $Z$ is exact in the limit that $N\rightarrow\infty$ but is not dependent on the value of $\Ne$.
The advantage of this extension is to include more flexibility
for obtaining convergence of the method in that a different number of beads
can be used for the nuclear and electronic parts of the Hamiltonian.

As in the usual derivation of path-integrals,
we insert identities of complete sets of nuclear position states as well as mapping position and momentum states.
These will become the 
ring-polymer bead positions,
$\mathbf{x}=\{\mat{x}_i\}$
and the corresponding
mapping variables,
$\mathbf{X}=\{\mat{X}_{i\alpha}\}$ and
$\mathbf{P}=\{\mat{P}_{i\alpha}\}$,
where
$\mat{X}_{i\alpha}=(X_{i\alpha 0},X_{i\alpha 1})$
and
$\mat{P}_{i\alpha}=(P_{i\alpha 0},P_{i\alpha 1})$
are vectors.
The nuclear-bead index is $i\in\{1,\dots,N\}$ and the mapping-variable index for each bead is $\alpha\in\{1,\dots,\Ne\}$,
whereas the third index	refers to the electronic state $n\in\{0,1\}$.

Using the imaginary-time free-particle propagator,
\begin{align}
\braket{\mat{x}_{i-1}|\eu{-\beta_N|\op{\mat{p}}|^2/2m}|\mat{x}_i} = \left(\frac{m}{2\pi\beta_N\hbar^2}\right)^{f/2} \eu{-m|\mat{x}_i-\mat{x}_{i-1}|^2/2\beta_N\hbar^2} \, ,
\end{align}
where $f$ is the number of nuclear degrees of freedom,
i.e. the dimension of the vector $\mat{x}_i$,
the partition function becomes
\begin{align}
	Z
	&\simeq \Tr\left[ \prod_{i=1}^N \int\rmd\mat{x}_i \, \eu{-\beta_N |\op{\mat{p}}|^2/2m} \ketbra{\mat{x}_i}{\mat{x}_i} \eu{-\beta_N U(\op{\mat{x}})} \right.
		\nonumber\\&\qquad\times
		\left. \prod_{\alpha=1}^\Ne \iint\rmd\mat{X}_{i\alpha}\rmd\mat{P}_{i\alpha} \, \eu{-\beta_N \mat{V}(\op{\mat{x}})/2\Ne} \mathcal{P} \ketbra{\mat{X}_{i\alpha}}{\mat{X}_{i\alpha}} \mathcal{P} \eu{-\beta_N \mat{V}(\op{\mat{x}})/2\Ne} \mathcal{P} \ketbra{\mat{P}_{i\alpha}}{\mat{P}_{i\alpha}} \mathcal{P} \right]
	\nonumber\\
	&= \iiint \tr\left[ \prod_{i=1}^N \braket{\mat{x}_{i-1}|\eu{-\beta_N|\op{\mat{p}}|^2/2m}|\mat{x}_i} \eu{-\beta_N U(\mat{x}_i)} \right.
		\nonumber\\&\qquad\times
		\left. \prod_{\alpha=1}^\Ne \eu{-\beta_N \mat{V}(\mat{x}_i)/2\Ne} \mathcal{P} \ketbra{\mat{X}_{i\alpha}}{\mat{X}_{i\alpha}} \mathcal{P} \eu{-\beta_N \mat{V}(\mat{x}_i)/2\Ne} \mathcal{P} \ketbra{\mat{P}_{i\alpha}}{\mat{P}_{i\alpha}} \mathcal{P} \right]
	\rmd\mathbf{x} \, \rmd\mathbf{X} \rmd\mathbf{P}
	\nonumber\\
	&= \left(\frac{m}{2\pi\beta_N\hbar^2}\right)^{\frac{Nf}{2}} \left(\frac{4}{\pi^2}\right)^{N\Ne} \iiint \eu{-\beta_N U_N(\mathbf{x})} 
		W_1 \eu{-|\mathbf{X}|^2 - |\mathbf{P}|^2}
	\rmd\mathbf{x} \, \rmd\mathbf{X} \rmd\mathbf{P} \, ,
\end{align}
where, using \eqn{wavefunctions},
\begin{align}
	W_1
	= \tr\left[ \prod_{i=1}^N \prod_{\alpha=1}^\Ne \mat{M}_i^{} \mat{X}_{i\alpha}^{} \mat{X}_{i\alpha}^\T \mat{M}_i^{} \mat{P}_{i\alpha}^{} \mat{P}_{i\alpha}^\T \right]
\end{align}
and we have defined the $2\times 2$ matrices
\begin{align}
	\mat{M}_i = \eu{-\beta_N\mat{V}(\mat{x}_i)/2\Ne} \, .
\end{align}
Note that we use the notation $\Tr[\cdot]$ for a full quantum-mechanical trace and $\tr[\cdot]$ for the trace of the matrix of electronic states only.
The standard ring-polymer potential \cite{TuckermanBook} is given by
\begin{equation}
	U_N(\mathbf{x}) = \sum_{i=1}^N \frac{m}{2\beta_N^2\hbar^2} |\mat{x}_i - \mat{x}_{i-1}|^2 + U(\mat{x}_i) \, ,
\end{equation}
where $i$ is considered as a cyclic variable such that $\mat{x}_0\equiv\mat{x}_N$.

For later convenience,
we define the ring-polymer momenta as $\mathbf{p}=\{\mat{p}_i\}$
and introduce them,
using the Gaussian integral identity,
into the expression for $Z$ to give
\begin{align}
	Z = \frac{1}{(2\pi\hbar)^{Nf}} \iiiint \rho_1(\mathbf{x},\mathbf{p},\mathbf{X},\mathbf{P}) \, \rmd \mathbf{x} \, \rmd\mathbf{p} \, \rmd\mathbf{X} \, \rmd\mathbf{P} \, ,
\end{align}
where the Boltzmann distribution in this representation is
\begin{gather}
	\rho_1(\mathbf{x},\mathbf{p},\mathbf{X},\mathbf{P}) = \left(\frac{4}{\pi^2}\right)^{N\Ne} W_1 \, \eu{-|\mathbf{X}|^2 - |\mathbf{P}|^2 -\beta_N[|\mathbf{p}|^2/2m+U_N(\mathbf{x})]} \label{rho1} \, .
\end{gather}

Following a similar approach,
we are able to derive a formula for the exact expression for quantum statistical quantities in terms of ring-polymer beads and mapping variables.
The average value of an operator, $\op{A}$, can be obtained in the $N\rightarrow\infty$ limit using
\begin{align}
	\label{CA}
	\frac{1}{Z}\Tr\big[\eu{-\beta\op{H}}\op{A}\big]
	= \braket{\bar{A}}_{\rho_1}
	\equiv \frac{1}{Z} \frac{1}{(2\pi\hbar)^{Nf}} \iiiint \rho_1 \bar{A} \, \rmd\mathbf{x} \, \rmd\mathbf{p} \, \rmd\mathbf{X} \, \rmd\mathbf{P} \, ,
\end{align}
where the form of $\bar{A}$ 
must, in principle, be derived depending on the form of $\op{A}$ such that \eqn{CA} is satisfied.
Note that $W_1$ is not in general positive definite
so in order to perform the average using a Monte Carlo method,
the sampling distribution should be the absolute value of $\rho_1$
and the results weighted by its sign:
\begin{align}
	\braket{\bar{A}}_{\rho_1} = \frac{\braket{\bar{A} \sgn\rho_1}_{|\rho_1|}}{\braket{\sgn\rho_1}_{|\rho_1|}} \, .
\end{align}

A common case of interest will be the expectation value of a function of position $A(\op{\mat{x}})$
which can be evaluated exactly using $\bar{A} = \frac{1}{N} \sum_{i=1}^N A(\mat{x}_i)$
as in the usual ring-polymer formulation \cite{RPMDcorrelation,Chandler+Wolynes1981,Parrinello1984Fcenter,TuckermanBook}.
For operators of electronic states, there are two alternative approaches, both of which give exact results in the $N\rightarrow\infty$ limit.
For instance, to obtain the expectation value of the operator $\op{A}=\ketbra{n}{m}$,
one option is to replace the trace in $W_1$ by the element which is the $m$th row and $n$th column of the matrix.
This can be formally represented using \eqn{CA} and 
\begin{subequations}
\label{Abar}
\begin{align}
	\bar{A} = \frac{P_{N\Ne n} [\mat{M}_1\mat{X}_{11}]_{m}}{\mat{P}_{N\Ne}^\T \mat{M}_1^{} \mat{X}_{11}^{}}
\end{align}
or equivalently, using the properties of cyclic permutation of the beads to make a more symmetric form,
\begin{align}
	\bar{A} = \frac{1}{N\Ne} \sum_{i=1}^N \sum_{\alpha=1}^\Ne \frac{P_{i(\alpha-1)n} [\mat{M}_i\mat{X}_{i\alpha}]_{m}}{\mat{P}_{i(\alpha-1)}^\T \mat{M}_i^{} \mat{X}_{i\alpha}^{}} \, .
\end{align}
\end{subequations}
According to the cyclic properties of the ring polymer,
when $\alpha-1=0$, the index $i(\alpha-1)$ is understood to mean $(i-1)\Ne$.

A second approach uses the mapping variable representation of the operator as given in \eqn{mapping}
to obtain the ring-polymer estimator
\begin{align}
	\label{Bbar}
	\bar{B}(\mathbf{X},\mathbf{P}) = \frac{1}{N \Ne} \sum_{i=1}^N \sum_{\alpha=1}^\Ne B(\mat{X}_{i\alpha},\mat{P}_{i\alpha}) \, ,
\end{align}
where $B(\mat{X}_{i\alpha},\mat{P}_{i\alpha})$ is 
obtained by replacing quantum operators in the mapping representation of $\op{B}$ by classical coordinates.
For instance the population operator $\op{B}=\ketbra{n}{n}$ is
represented by \eqn{Bbar} with
$B(\mat{X}_{i\alpha},\mat{P}_{i\alpha}) = \half (X_{i\alpha n}^2 + P_{i\alpha n}^2 - 1)$.
The proof that average values of $\bar{B}$ defined in this way
tend to the exact quantum statistics in the $N\rightarrow\infty$ limit
is given at the end of \secref{dynamics}
and
another example of one of these operators is given in \secref{vibronic}.

\subsection{Nonadiabatic ring-polymer dynamics}
\label{sec:dynamics}

As we have shown, quantum statistics can be computed exactly from the partition function and derivatives of it
using the ring-polymer mapping formulation.
However, there is no known method for efficiently obtaining exact dynamical information in this way.
Instead we employ an extension of the approximate RPMD ansatz
\cite{RPMDcorrelation,Habershon2013RPMDreview,Hele2015RPMD} to treat systems with mapping variables \cite{mapping}.

To obtain the dynamical equations of motion,
we append the standard RPMD Hamiltonian, which includes spring terms between the nuclear beads,
with a sum over the mapping representation of the diabatic potential for each set of variables.
This gives the NRPMD Hamiltonian
\begin{align}
	\label{HNRPMD}
	\mathcal{H}_{N,\Ne} %
	= \frac{|\mathbf{p}|^2}{2m} + U_N(\mathbf{x})
	+ \frac{1}{\Ne} \sum_{i=1}^N 
		\sum_{\alpha=1}^\Ne
		\thalf
		\left[ \mat{X}_{i\alpha}^\T\mat{V}(\mat{x}_i)\mat{X}_{i\alpha}^{} + \mat{P}_{i\alpha}^\T\mat{V}(\mat{x}_i)\mat{P}_{i\alpha}^{} - \tr\mat{V}(\mat{x}_i) \right],
\end{align}
which reduces to the classical mapping Hamiltonian for $N=\Ne=1$.
The canonical variables,
$\{\mat{x}_i\}$, $\{\mat{p}_i\}$, $\{\sqrt\frac{\hbar}{\Ne}X_{i\alpha n}\}$ and $\{\sqrt\frac{\hbar}{\Ne}P_{i\alpha n}\}$,
are used to obtain Hamilton's equations of motion
\cite{Goldstein}:
\begin{subequations}
\label{dynamics}
\begin{align}
	\dot{\mat{x}}_i &= \frac{\mat{p}_i}{m}
\\
	\dot{\mat{p}}_i &=
	- \pder{U_N}{\mat{x}_i}
	- \frac{1}{\Ne} \sum_{\alpha=1}^\Ne
		\half \left[ \mat{X}_{i\alpha}^\T\pder{\mat{V}}{\mat{x}_i}\mat{X}_{i\alpha}^{} + \mat{P}_{i\alpha}^\T\pder{\mat{V}}{\mat{x}_i}\mat{P}_{i\alpha}^{} - \tr\pder{\mat{V}}{\mat{x}_i} \right]
\\
	\dot{\mat{X}}_{i\alpha} &= \frac{1}{\hbar} \mat{V}(\mat{x}_i) \mat{P}_{i\alpha}
\\
	\dot{\mat{P}}_{i\alpha} &= -\frac{1}{\hbar} \mat{V}(\mat{x}_i) \mat{X}_{i\alpha} \, .
\end{align}
\end{subequations}

Using Liouville operators \cite{TuckermanBook},
a symplectic integrator with
exact harmonic solution of mapping variables
can be derived 
to give
\begin{align}
	\eu{\iu \mathcal{L} \delta t} \simeq \eu{\frac{\iu}{2} \mathcal{L}_{\mathbf{XP}} \delta t} \eu{\frac{\iu}{2} \mathcal{L}_{\mathbf{p}} \delta t} \eu{\iu \mathcal{L}_{\mathbf{x}} \delta t} \eu{\frac{\iu}{2} \mathcal{L}_{\mathbf{p}} \delta t} \eu{\frac{\iu}{2} \mathcal{L}_{\mathbf{XP}} \delta t} \, ,
\end{align}
where the approximation is valid in the limit $\delta t\rightarrow0$,
\begin{align}
	\mathcal{L} &= \mathcal{L}_{\mathbf{x}} + \mathcal{L}_{\mathbf{p}} + \mathcal{L}_{\mathbf{XP}}
\end{align}
and
\begin{subequations}
\begin{align}
	\iu \mathcal{L}_{\mathbf{x}} &= \sum_{i=1}^N \dot{\mat{x}}_i \cdot \pder{}{\mat{x}_i}
	\\
	\iu \mathcal{L}_{\mathbf{p}} &= \sum_{i=1}^N \dot{\mat{p}}_i \cdot \pder{}{\mat{p}_i}
	\\
	\iu\mathcal{L}_{\mathbf{XP}} &= \sum_{i=1}^N \sum_{\alpha=1}^\Ne \mat{P}_{i\alpha}^\T \mat{V}(\mat{x}_i) \pder{}{\mat{X}_{i\alpha}} - \mat{X}_{i\alpha}^\T \mat{V}(\mat{x}_i) \pder{}{\mat{P}_{i\alpha}} \, .
\end{align}
\end{subequations}

These operators have the following effect on variables:
\begin{subequations}
\begin{align}
	\eu{\iu \mathcal{L}_{\mathbf{x}} \delta t} \mat{x}_i &= \mat{x}_i + \dot{\mat{x}}_i \delta t
	\\
	\eu{\frac{\iu}{2} \mathcal{L}_{\mathbf{p}} \delta t} \mat{p}_i &= \mat{p}_i + \thalf \dot{\mat{p}}_i \delta t
	\\
		\eu{\frac{\iu}{2}\mathcal{L}_{\mathbf{XP}} \delta t} \begin{pmatrix} \mat{X}_{i\alpha} \\[1ex] \mat{P}_{i\alpha} \end{pmatrix}
		&= \begin{pmatrix} \cos\frac{\mat{V}(\mat{x}_i)\delta t}{2\hbar} & \sin\frac{\mat{V}(\mat{x}_i)\delta t}{2\hbar} \\[1ex]
							 - \sin\frac{\mat{V}(\mat{x}_i)\delta t}{2\hbar} & \cos\frac{\mat{V}(\mat{x}_i)\delta t}{2\hbar} \end{pmatrix}
		\begin{pmatrix} \mat{X}_{i\alpha} \\[1ex] \mat{P}_{i\alpha} \end{pmatrix}
\end{align}
\end{subequations}
and otherwise leave them unchanged.
Performing the mapping-variable update analytically avoids limiting the integration time step to describe the fast mapping-variable oscillations.
An alternative approach avoids the stiff equations of motion by transforming to a new set of coordinates \cite{Wang1999mapping}.
As is common in path-integral molecular dynamics simulations,
it is also possible to perform a similar trick on the ring-polymer springs
by separating
the free ring-polymer normal modes 
from the Liouville operator
and treating their time evolution exactly \cite{TuckermanBook}.

Solving these equations numerically allows trajectories to be computed and hence the time dependence of operators.
Note that it is expensive to evaluate each nuclear bead as the potential matrix needs to be computed at each nuclear configuration.
However, each additional mapping variable requires no significant computation to obtain a trajectory.%
\footnote{Although the sign problem is still a cause of reduced efficiency when increasing $\Ne$.}
Therefore by allowing $\Ne$ to be large but keeping $N$, the number of beads, small,
we are able to approach the convergence limits more efficiently.

Again choosing $\Ne=1$ recovers the original NRPMD formulation \cite{mapping}.
Results from the extended method are not expected to deviate
from those of the original method in the $N\rightarrow\infty$ limit.
This is a consequence of the strong spring terms which force neighbouring ring-polymer beads to bunch up
and
due to an adiabatic separation of the higher ring-polymer modes, these groups act like single ring-polymer beads with many mapping variables.
The main advantage of the new extended methods is in the improved ability to control the convergence of $N$ and $\Ne$ separately.
This flexibility potentially makes the method more efficient and additionally makes our following mathematical analyses simpler.

A new nonadiabatic classical trajectory method also presents itself 
if we consider $N=1$ but allow $\Ne>1$.
This generalizes the classical mapping approach to many mapping variables but ignores quantum nuclear effects.
This approach will benefit from the improved dynamical properties discovered in \secref{accuracy}
but is simpler than the full NRPMD approach as it does not involve a ring polymer.
It may be a useful approach for studying nonadiabatic effects in systems where the nuclear masses are large
and thus show limited quantum effects.
Note however that nuclear tunnelling effects often accompany nonadiabatic transitions even for large masses at high temperatures \cite{GoldenGreens,GoldenRPI}.

The mapping approach (and therefore NRPMD) gives equivalent dynamics whether formulated in the adiabatic or diabatic representation
\cite{Mueller1998mapping,Sun1997mapping}.
This is an important fact, especially when it is noted that other
approximate nonadiabatic dynamics methods such as Ehrenfest and surface hopping \cite{Tully1990hopping} are not independent of the choice of representation
\cite{Stock2005nonadiabatic}.

The RPMD ansatz is designed to obtain approximations
to the Kubo-transformed correlation function, \cite{RPMDcorrelation}
defined by
\begin{equation}
	\label{CAB}
	\tilde{C}_{AB}(t) = \frac{1}{Z\beta} \int_0^\beta \Tr\left[\eu{-(\beta-\lambda)\op{H}} \op{A} \, \eu{-\lambda\op{H}} \eu{\iu\op{H}t/\hbar} \op{B} \, \eu{-\iu\op{H}t/\hbar}\right] \rmd\lambda
\end{equation}
for operators $\op{A}$ and $\op{B}$ of nuclear position or electronic states.
Following the RPMD ansatz, with the generalization to nonadiabatic dynamics described above,
this correlation function is approximated by
\cite{mapping},
\begin{equation}
	\label{Capprox}
	\tilde{C}_{AB}(t) \approx \braket{\bar{A}(\mathbf{x}_0,\mathbf{X}_0,\mathbf{P}_0) \, \bar{B}(\mathbf{x}_t,\mathbf{X}_t,\mathbf{P}_t)}_{\rho_1},
\end{equation}
where the initial values (with subscript 0) are obtained from the distribution $\rho_1$
and trajectories propagated to time $t$ according to the dynamics defined in \eqn{dynamics}.

The approximate correlation functions were tested against the exact quantum results in \Ref{mapping}
for a model system
where they compared favourably for a range of parameters.
In each case it was seen that increasing the value of $N$ improved the results, not only at short times but also for longer times.

When dealing with electronic-state operators, it is important to use the first type, \eqn{Abar}, for $\op{A}$ and the second type, \eqn{Bbar}, for $\op{B}$.
This is because the derivation of the first type is only valid at $t=0$ and thus cannot be used for $\op{B}$ which needs to be computed at all times.
One might expect that the second type can be used for both operators, but this is can lead to poor convergence in certain situations
as we shall show in the following example.

The calculation of the NRPMD approximation of $\tilde{C}_{AB}(0)$ for $\op{A}=\op{B}=\ketbra{n}{n}$ can be carried out as follows.
The distribution, $\rho_1$, contains products of terms of the type
\begin{align}
	\mat{G}_{i\alpha} &= 
		\frac{4}{\pi^2}
		\,
		\mat{M}_i^{} \mat{X}_{i\alpha}^{} \mat{X}_{i\alpha}^\T \mat{M}_i^{} \mat{P}_{i\alpha}^{} \mat{P}_{i\alpha}^\T
		\,\eu{-|\mat{X}_{i\alpha}|^2-|\mat{P}_{i\alpha}|^2} \, ,
\end{align}
which by construction are normalized
such that
$%
	\iint \mat{G}_{i\alpha} 
	\,\rmd\mat{X}_{i\alpha} \rmd\mat{P}_{i\alpha}
	= \mat{M}_i^2
$. %
Consider the effect of multiplying this distribution
by the $n$th diabatic state population estimator, $\bar{B}$.
This gives integrals of the type
\begin{align}
	\mat{B}_{i\alpha} &=
	\iint
	\mat{G}_{i\alpha}
	\thalf (X_{i\alpha n}^2 + P_{i\alpha n}^2 - 1)
	\,\rmd\mat{X}_{i\alpha} \rmd\mat{P}_{i\alpha}
	\nonumber\\
	&=
	\thalf \big( \mat{M}_i \ketbra{n}{n} \mat{M}_i + \mat{M}_i \mat{M}_i \ketbra{n}{n} \big) \, ,
	\label{Bia}
\end{align}
and thus the estimator has the same effect as inserting a projection onto the corresponding electronic state.
Therefore a path-integral calculation of $\braket{\bar{B}}_{\rho_1}$ gives,
in the limit of $N\rightarrow\infty$, the exact result, $\Tr\big[\eu{-\beta\op{H}}\ketbra{n}{n}\big]$,
for the population of the $n$th state. %

The short-time limit of the NRPMD correlation function can also be evaluated explicitly as it does not involve any dynamics.
It is
\begin{align}
	\braket{\bar{A} \bar{B}}_{\rho_1}
	=
	\frac{1}{Z} \frac{1}{(2\pi\hbar)^{Nf}}
	\iint
	\frac{1}{N\Ne} \sum_{i=1}^N \sum_{\alpha=1}^\Ne
	\braket{n|\mat{M}_1^2\cdots\mat{B}_{i\alpha}\cdots\mat{M}_N^2|n}
	\eu{-\beta[|\mathbf{p}|^2/2m + U_N(\mathbf{x})]}
	\, \rmd\mathbf{x} \, \rmd\mathbf{p} \, ,
\end{align}
which includes a product over matrices $\mat{M}_i^2$ for each set of mapping variables where one, corresponding to $i\alpha$, is replaced by $\mat{B}_{i\alpha}$.
This result
tends to the right answer, $\tilde{C}_{AB}(0)$, as $N\Ne\rightarrow\infty$ %
as the sum describes the Kubo transform.

If we were to have evaluated the short-time limit of this correlation function
using
$\braket{\bar{B} \bar{B}}_{\rho_1}$,
we would have found an expression which 
includes terms where the estimator occurs twice for the same set of of mapping variables.
This integral does not take on such a simple expression as in \eqn{Bia}
and the average is only equal to the exact expression in the limit of 
very many mapping variables such that these terms are drowned out.
There is no problem with applying $\bar{A}$ and $\bar{B}$ to the same set and so,
to improve convergence, we use the $\braket{\bar{A}\bar{B}}_{\rho_1}$ form.

It is well-known that the classical mapping dynamics is able to describe Rabi oscillations exactly \cite{Wang1999mapping}
in a system where the potential-energy matrix does not depend on the nuclear positions, $\mat{V}(\mat{x}) = \mat{V}$.
In this case, the NRPMD equations of motion reduce simply to a set of $N\Ne$ uncoupled sets of classical mapping variables
each of which oscillate with the Rabi frequency.
As we have shown above that the statistics are correct in the $\Ne\rightarrow\infty$ limit,
any Kubo-transformed correlation function of this uncoupled system will also be calculated exactly using the NRPMD approach.

\section{Analysis of the method}
\label{sec:analysis}

In this section we analyse both the efficiency and accuracy of using the NRPMD approach to approximate quantum dynamics
and seek to improve its efficiency
while showing that it is more accurate than formerly believed.

A computational method based on classical trajectories is generally considered to be very efficient,
especially when compared to exact quantum dynamics methods.
However, in this case,
there is a particular difficulty associated with sampling the initial conditions for the trajectories,
which is that the distribution $\rho_1$ is not positive definite
and can lead to poor statistics due to cancellation of positive and negative weights.
We call this the sign problem and, unless dealt with, it will limit the
number of mapping variables which can be used in a practical simulation.
Here we study the cause of the sign problem
and find a simple way to increase the efficiency without affecting the accuracy of the results.

A potential flaw with the dynamics is that an ensemble of trajectories
does not explicitly
preserve the initial distribution.
This results in a failure of detailed balance
and allows an unphysical leakage of zero-point energy from the mapping modes \cite{Stock2005nonadiabatic}.
However, this process does not happen immediately and correlation functions may still be approximated fairly accurately at least for fairly short times
\cite{mapping}.
The dynamics can also be found to vary depending on the choice of separation into $U(\mat{x})$ and $\mat{V}(\mat{x})$ parts.
This is a potential problem for the accuracy and predictive power of the method but as we shall show in this section,
may be alleviated by increasing the number, $\Ne$, of mapping variables per bead.

\subsection{Efficiency}

The analysis of the method is more easily carried out
after transforming the mapping variables
to action-angle coordinates,
$\mathbf{J}=\{J_{i\alpha n}\}$ and $\mathbf{\theta}=\{\theta_{i\alpha n}\}$,
where
\begin{align*}
	X_{i\alpha n} &= \sqrt{2J_{i\alpha n}+1} \sin{\theta_{i\alpha n}} &
	P_{i\alpha n} &= \sqrt{2J_{i\alpha n}+1} \cos{\theta_{i\alpha n}} \, .
\end{align*}
The initial angle variables are chosen in the ranges
$-\pi \le \theta_{i\alpha n} < \pi$.
The actions,
$-\thalf \le J_{i\alpha n} < \infty$,
play an important role in the NRPMD theory as the population estimators of the diabatic states for a particular bead and mapping variable set.
The angles for the two states are then transformed to sum and difference coordinates,
$\mathbf{\Theta}=\{\Theta_{i\alpha}\}$ and $\mathbf{\vartheta}=\{\vartheta_{i\alpha}\}$,
where
\begin{align*}
	\Theta_{i\alpha} &= \thalf (\theta_{i\alpha 0} + \theta_{i\alpha 1}) &
	\vartheta_{i\alpha} &= \theta_{i\alpha 1} - \theta_{i\alpha 0}
\end{align*}
such that 
$-2\pi \le \vartheta_{i\alpha} < 2\pi$
and
$-\pi + |\vartheta_{i\alpha}|/2 \le \Theta_{i\alpha} < \pi - |\vartheta_{i\alpha}|/2$.

Note that the transformed NRPMD Hamiltonian,
\begin{align}
	\label{HJtheta}
	\mathcal{H}_{N,\Ne}
	= \frac{|\mathbf{p}|^2}{2m} + U_N(\mathbf{x})
	+ \frac{1}{\Ne} \sum_{i=1}^N 
		\sum_{\alpha=1}^\Ne
		\left[
		\sum_{n=0}^1
		J_{i\alpha n} V_n(\mat{x}_i)
	+ \Delta(\mat{x}_i) \sqrt{2J_{i\alpha 0}+1}\sqrt{2J_{i\alpha 1}+1} \cos{\vartheta_{i\alpha}}
	\right],
\end{align}
does not
include the variables $\Theta_{i\alpha}$.
These are therefore known as cyclic variables \cite{Goldstein}
and it is because of this that the total electronic population, $J_{i\alpha 0}+J_{i\alpha 1}$, is constant.
Assuming the observables are also invariant to this,
we can perform the integral of the distribution with respect to $\Theta_{i\alpha}$ explicitly
to improve the efficiency of the NRPMD method without affecting its results.

The new distribution is %
\begin{align}
	W_2(\mathbf{x},\mathbf{J},\mathbf{\vartheta})
	&= 
	\int
	W_1(\mathbf{x},\mathbf{J},\mathbf{\theta}) 
	\,\rmd\mathbf{\Theta}
\\
	&= \tr\left[ \prod_{i=1}^N %
	\prod_{\alpha=1}^\Ne \mat{\Phi}_{i\alpha}
	\right],
	\label{W2}
\end{align}
where elements of the $2\times2$ matrices $\mat{\Phi}_{i\alpha}$ are
\begin{align}
	[\mat{\Phi}_{i\alpha}]_{nm} = \sum_{q,r,s=0}^1 [\mat{M}_i]_{nq} [\mat{M}_i]_{rs} [\mat{T}_{i\alpha}]_{qrsm} 
\end{align}
and
\begin{align}
	\label{Ti}
	[\mat{T}_{i\alpha}]_{qrsm} =
	\int_{-\pi+|\vartheta_{i\alpha}|/2}^{\pi-|\vartheta_{i\alpha}|/2} X_{i\alpha q} X_{i\alpha r} P_{i\alpha s} P_{i\alpha m} \, \rmd \Theta_{i\alpha} \, .
\end{align}
Analytic results for these integrals are given in \Appendix.
The new, more efficient approach is defined by \eqn{Capprox}
but using the distribution $\rho_2$ instead of $\rho_1$,
where
\begin{align}
	\rho_2(\mathbf{x},\mathbf{p},\mathbf{J},\mathbf{\vartheta})
	&= \int \rho_1(\mathbf{x},\mathbf{p},\mathbf{X},\mathbf{P}) \, \rmd \mathbf{\Theta}
	\\
	&=
	\left(\frac{4}{\pi^2}\right)^{N\Ne}
	W_2 \, \eu{-\sum_{i\alpha n} (2J_{i\alpha n}+1) -\beta_N[|\mathbf{p}|^2/2m+U_N(\mathbf{x})]} \label{rho2} \, .
\end{align}
Note that this distribution is only valid within the integration ranges
of the initial variables
and should be zero otherwise.
The action-angle coordinates chosen from this distribution can be easily transformed back to mapping variables
in order to perform the dynamics in the simplest representation.
It is necessary to specify a value of $\Theta_{i\alpha}$ for this but the choice does not affect the results.

We can perform a simple test to show how this procedure has greatly improved the sign problem.
Taking the high-temperature limit of $\mat{M}_i$ as the identity matrix,
we compute $\sigma_d = \braket{\sgn{W_d}}_{|\rho_d|}$,
where $d$ refers either to the old ($d=1$) or new ($d=2$) distribution.
This is a measure of the efficiency of the method
such that when $\sigma_d=1$, the whole distribution is positive definite
and as $\sigma_d$ tends to 0, the sign problem worsens.
Results are shown in Table~\ref{table:signs} for the old and new distributions.

\begin{table}
\centering
\caption{The variation in the average sign of the initial distribution with the total number of mapping variables,
using either the old factor $W_1$ or the new distribution with $W_2$.
The system is described in the main text.}
\label{table:signs}
\begin{tabular}{lll}
\hline
	$N\Ne$ & $\sigma_1$ & $\sigma_2$ \\
\hline
	1 & 1 & 1 \\
	2 & 0.74 & 0.86 \\
	4 & 0.29 & 0.47 \\
	8 & 0.04 & 0.11 \\
\hline
\end{tabular}
\end{table}

It is seen that the new distribution greatly reduces the number of
trajectories weighted by a negative number which will enhance sampling efficiency.
In either case, the $N\Ne=1$ distribution is positive definite,
which is a general result true for any $\mat{M}_i$.
In the worst case, with $W_1$ for $N\Ne=8$,
96\% of all initial conditions are cancelled out by another with the opposite sign.
In the new distribution, there are still 11\% of trajectories contributing without being cancelled out.
This is thus almost a three-fold improvement in the efficiency.
That is, one expects to need to run three times fewer trajectories to obtain the same statistical error.

There is no sign problem at all, with either the old or new distributions for the limiting cases of
\begin{align*}
	\mat{M}_i &= \begin{pmatrix} 1 & 1 \\ 1 & 1 \end{pmatrix}
	& \text{or} &&
	\mat{M}_i &= \begin{pmatrix} 1 & 0 \\ 0 & 0 \end{pmatrix}.
\end{align*}
This is
because, in these cases,
the mapping variables, $X_{i\alpha n}$ and $P_{i\alpha n}$,
always appear in pairs and thus multiply to give positive values only.

\subsection{Accuracy}
\label{sec:accuracy}

In standard RPMD, the dynamics of the individual beads are not themselves physically significant
and it is only averages over the beads which are used to probe the quantum dynamics of the system.
The same is true of NRPMD dynamics for which the information from the mapping variables should only be interpreted as
information about the electronic states after it has been averaged.
In this subsection we analyse the properties of averages over 
mapping variables and show that certain problems associated with the classical mapping approach
are resolved in the case of NRPMD.

According to the exact quantum dynamics obtained from the Hamiltonian, \eqn{Hdiabatic},
the results should not change
if a function $f(\mat{x})$ is added to $U(\mat{x})$ and subtracted from the diagonal elements of $\mat{V}(\mat{x})$. 
This is because of the identity of the total electronic population operator, $\sum_{n=0}^1 \ketbra{n}{n} = 1$.
However, if we perform the same trick with the NRPMD Hamiltonian, \eqn{HJtheta},
we find an extra term \mbox{$\sum_{i=1}^N f(\mat{x_i}) \left[1 - \sum_{n=0}^1 \bar{J}_{in}\right]$},
where \mbox{$\bar{J}_{in}=\frac{1}{\Ne}\sum_{\alpha=1}^\Ne J_{i\alpha n}$}.
This changes the forces on ring-polymer beads and makes our results non-unique
as they depend on the choice of $f(\mat{x})$.
We have already identified $J_{i\alpha n}$
as an $n$th-state population estimator 
and thus $\bar{J}_{in}$ gives the average population of bead $i$.
Only if the total population of each bead $\sum_{n=0}^1 \bar{J}_{in}$ is exactly 1 will the forces be uniquely defined.
We shall show that in the limit of $\Ne\rightarrow\infty$,
this is indeed the case for the NRPMD approach.

Another related issue with the classical mapping dynamics is that
they can suffer from unphysical behaviour
if the value of $\bar{J}_{in}$ is negative.
In this case,
the direction of the forces is reversed and
trajectories may end up following an inverted potential-energy surface
which leads to unphysical results \cite{Stock2005nonadiabatic,Bonella2001mapping1}.
We shall also investigate the behaviour of the NRPMD method with respect to this issue.

The NRPMD distribution, $\rho_2$, can be written as a sum over terms
by expanding the matrix multiplications of \eqn{W2}.
According to the integrals given in \Appendix,
each term
carries a weight proportional to \mbox{$(2J_{i\alpha n}+1)^\gamma \, \eu{-(2J_{i\alpha n}+1)}$}
for some value of $\gamma$ which is given explicitly by the formulae
as one of $\{0,\half,1,\tfrac{3}{2},2\}$.
A Monte Carlo implementation of the integral over $\rho_2$
would select
initial variables, $J_{i\alpha n}$,
randomly from the Gamma distribution
\begin{align}
	J_{i\alpha n} \sim \frac{2}{\Gamma(1+\gamma)} (2J_{i\alpha n}+1)^\gamma \, \eu{-(2J_{i\alpha n}+1)} \, ,
\end{align}
which has mean $\gamma/2$ and variance $(1+\gamma)/4$.
According to the central-limit theorem,
the average of many of these variables, $\bar{J}_{in}$,
will be distributed by the normal distribution
with mean $\gamma/2$ and variance $(1+\gamma)/4\Ne$,
\begin{align}
	\bar{J}_{in}
	& \sim
	\sqrt\frac{2\Ne}{\pi(1+\gamma)} \, \eu{-2\Ne(\bar{J}_{in}-\gamma/2)^2/(1+\gamma)} \, .
\end{align}
Therefore as $\Ne\rightarrow\infty$, 
the width decreases to a delta function
and \mbox{$\bar{J}_{in} \sim \delta(\bar{J}_{in}-\gamma/2)$} will always be found to be $\gamma/2$.
It cannot therefore be negative and thus the problem of 
the dynamics following an inverted potential has been avoided,
at least at the start of the simulation.

It so happens that for the Boltzmann distribution,
the $\gamma$ value for $J_{i\alpha 0}$ and the $\gamma$ value for $J_{i\alpha 1}$ always sum to 2.
This is because each term in \eqn{Ti} has exactly four mapping variables,
each of which contributes a factor of a half.
Therefore the total average electronic population is found to be $\sum_{n=0}^1 \bar{J}_{in} = 1$,
and the choice of separation of potential into state-dependent, $\mat{V}(\mat{x})$,
and state independent, $U(\mat{x})$, parts is unimportant.
Because $\sum_{n=0}^1 \bar{J}_{in}$ is constant, a consequence of the absence of $\Theta_{i\alpha}$ from $\mathcal{H}_{N,\Ne}$,
this identity holds for the whole length of the trajectory
and thus we have proved that the NRPMD dynamics are uniquely defined.

Nonetheless, it may help convergence to choose the separation wisely.
Depending on the system under study,
an intelligent choice may be defined 
such that either $\tr\mat{V}(\mat{x})=0$ or that the lowest eigenvalue of $\mat{V}(\mat{x})$ is 0 \cite{mapping}.

In summary,
the NRPMD approach is not unique with respect to the choice of the potential separation
with a small number of mapping variables.
However, we have shown that in the limit of an infinite number of mapping variables,
the method becomes unique.
This arises automatically from the ring-polymer distribution
and does not need to be imposed as was done for the 
classical mapping approach
\cite{Mueller1998mapping,Stock2005nonadiabatic,Bonella2003mapping}
where the value of the total population is fixed to 1 in order to remove this problem.
Such a modification to our approach would have destroyed the exactness of the distribution at initial times.

We have also shown that the initial distribution forbids negative values of $\bar{J}_{in}$
and thus avoids following inverted potentials.
However, if zero-point energy leakage occurs \cite{Stock2005nonadiabatic},
the problem may still be found at longer times along the trajectory.
Whether the zero-point energy leakage problem is resolved in the limit of $\Ne\rightarrow\infty$ is not yet known 
and will be the subject of future work.

As an extension to these results,
it is also possible to show 
that for uncoupled systems where $\Delta(\mat{x})=0$, such as is treated in \secref{vibronic},
the initial distribution
of electronic populations and nuclear configurations
is conserved at all times by the ensemble of trajectories.
This is also true of the large $\Delta$ limit such that the Born-Oppenheimer approximation is valid,
and one can therefore speculate that for nonadiabatic systems between these two limits
the Boltzmann distribution may be conserved for $\Ne\rightarrow\infty$.
Testing this conjecture will also be the subject of future work.

\section{Vibronic spectra}
\label{sec:vibronic}

There is significant interest in computing vibronic spectra in complex molecular systems
using simulations based on classical and semiclassical trajectories \cite{
Stock1993spectroscopy,
Thoss2004review,Stock2005nonadiabatic,Egorov1998vibronic,Rabani1998vibronic, Wang2001Filinov, Wang2004absorption}.
In many molecular systems of interest
there exists a large energy gap between the ground and first excited electronic state.
We therefore assume that the ground and first excited electronic states are only coupled through interaction with the light field.
The simplest model for this problem
uses the Hamiltonian of an uncoupled two-state system as \eqn{Hdiabatic} with $\Delta(\mat{x})=0$.
Note that it has also been assumed that there are no other excited states which can interact.
For certain systems, a more comprehensive treatment will be required,
which for instance couples a second excited state to the first, e.g.\ through a conical intersection \cite{ConicalIntersections1}.
The mapping approach can be used to describe the nonadiabatic dynamics of these systems \cite{Stock1993spectroscopy,Uspenskiy2006spectra}
and thus the NRPMD approach will also be applicable.
Here, however, we shall only consider the simpler situation of the simulation of Franck-Condon spectra as a first step towards a more general treatment.

The observable of interest is the transition dipole moment,
\begin{align}
	\op{\mu} = \ketbra{0}{1} + \ketbra{1}{0}.
\end{align}
Here we have tacitly made the Condon approximation
and assume that it is a constant of the nuclear coordinates.
As we are only interested in the relative absorbance, 
we have set the magnitude of this operator to 1 and treat the dipole function and hence the correlation functions as dimensionless.
This assumption could be removed simply by multiplying the operator by the relevant function of $\op{\mat{x}}$.

The vibronic spectrum can be calculated
by simulating the dynamics to obtain
the dipole-dipole correlation function,
\begin{align}
	C_{\mu\mu}(t) &= \Tr\left[ \eu{-\beta\op{H}} \op{\mu} \, \eu{\iu\op{H}t/\hbar} \op{\mu} \, \eu{-\iu\op{H}/\hbar} \right],
\end{align}
from which the absorbance spectrum is given by
\begin{align}
	\label{spectrum}
	I(\omega)
	&= \frac{1}{2\pi} \int_{-\infty}^\infty C_{\mu\mu}(t) \, \eu{-\iu\omega t} \, \rmd t \, .
\end{align}
Note that this correlation function differs from the Kubo-transformed version, \eqn{CAB},
although they are related through their Fourier transforms \cite{RPMDcorrelation}:
\begin{align}
	\label{Kubo}
	\int_{-\infty}^\infty C_{\mu\mu}(t) \, \eu{-\iu\omega t} \, \rmd t
	=
	\frac{\beta\hbar\omega}{1-\eu{-\beta\hbar\omega}} 
	\int_{-\infty}^\infty \tilde{C}_{\mu\mu}(t) \, \eu{-\iu\omega t} \, \rmd t \, .
\end{align}

Egorov, Rabani and Berne suggested classical approaches
for approximating the correlation function and hence the vibronic spectrum
\cite{Egorov1998vibronic,Rabani1998vibronic}.
A number of different classical approximations are defined by \cite{MukamelBook}
\begin{align}
	\label{Ccl}
	C_\text{cl}(t) &= \Braket{\exp{\frac{\iu}{\hbar}\int_0^t \big[V_1(\mat{x}(t)) - V_0(\mat{x}(t))\big] \, \rmd t} }_{\rho_0} \, ,
\end{align}
where each case, the initial conditions are sampled from the ground electronic state, 
\begin{align}
	\rho_0 = \eu{-\beta\left[|\mat{p}|^2/2m + U(\mat{x}) + V_0(\mat{x})\right]} \, ,
\end{align}
but the dynamics differ.
In one method, called DCL, the classical dynamics are performed on the ground-state potential-energy surface, $U(\mat{x})+V_0(\mat{x})$,
whereas for ACL the effective surface is $U(\mat{x})+\thalf\big(V_0(\mat{x})+V_1(\mat{x})\big)$.
The two approaches are not equivalent,
and although ACL seems better in the example below,
it is not in general obvious which should be preferred in different situations \cite{Egorov1998vibronic,Rabani1998vibronic}.
An even simpler and more approximate approach, known as SCL, performs no dynamics at all such that $\mat{x}(t)=\mat{x}(0)$.

Note that the classical approaches make an additional assumption that the electronic energy gap is much greater than $k_\mathrm{B}T$
such that all trajectories are initialized in the ground state.
In the majority of cases, this is an excellent approximation.

We test the various approximations for the vibronic spectra on a simple one-dimensional harmonic system,
for which the exact quantum result is easily obtainable.
The potentials are defined as
$U(x) = \thalf m \omega_0^2 x^2$,
$V_0(x) = \kappa x$, and
$V_1(x) = \epsilon - \kappa x$.

For this harmonic system, the exact and classical correlation functions
can be obtained analytically.
They are
\begin{align}
	C_{\mu\mu}(t) &= \exp\left[\frac{\iu}{\hbar}\epsilon t + \frac{2\iu\kappa^2}{\hbar m\omega_0^3}\sin{\omega_0 t} - \frac{4\kappa^2}{\hbar m\omega_0^3} \frac{\sin^2{\half\omega_0 t}}{\tanh\half\beta\hbar\omega_0}\right]
	\\
	C_\text{ACL}(t) &= \exp\left[\frac{\iu}{\hbar}\epsilon t + \frac{2\iu\kappa^2}{\hbar m\omega_0^3}\sin{\omega_0 t} - \frac{2\kappa^2}{\beta \hbar^2 m\omega_0^4} \sin^2{\thalf\omega_0 t}\right]
	\\
	C_\text{DCL}(t) &= \exp\left[\frac{\iu}{\hbar}\epsilon t + \frac{2\iu\kappa^2}{\hbar m\omega_0^2}t - \frac{2\kappa^2}{\beta \hbar^2 m\omega_0^4} \sin^2{\thalf \omega_0 t}\right]
	\\
	C_\text{SCL}(t) &= \exp\left[\frac{\iu}{\hbar}\epsilon t + \frac{2\iu\kappa^2}{\hbar m\omega_0^2}t - \frac{\kappa^2}{2\beta \hbar^2 m\omega_0^2}t^2 \right].
\end{align}
In the exact case, as well as for the classical approximations,
we have again assumed that $\beta\epsilon\gg1$ such that the initial population of the excited state is 0.

It is easy to see how the correlation functions compare for this harmonic case.
The classical approximations assume that $\beta\hbar\omega_0\ll1$.
This is the only approximation made by ACL for this system.
DCL additionally makes an approximation to the phase of the correlation function
and SCL further approximates the decay by a Gaussian.
DCL and SCL are thus examples of short-time approximations as they are only valid in the limit $t\rightarrow0$.
Both the exact and ACL forms are periodic in $t$ and therefore give rise to a discrete spectrum.
DCL also has a discrete spectrum but at frequencies shifted slightly by the second term, unless $2\kappa^2/\hbar m \omega_0^3$ happens to be an integer.
The static approximation, SCL,
gives rise to a continuous spectrum
exhibiting inhomogeneous line broadening \cite{MukamelBook}.

\begin{figure}
	\centering
	\includegraphics{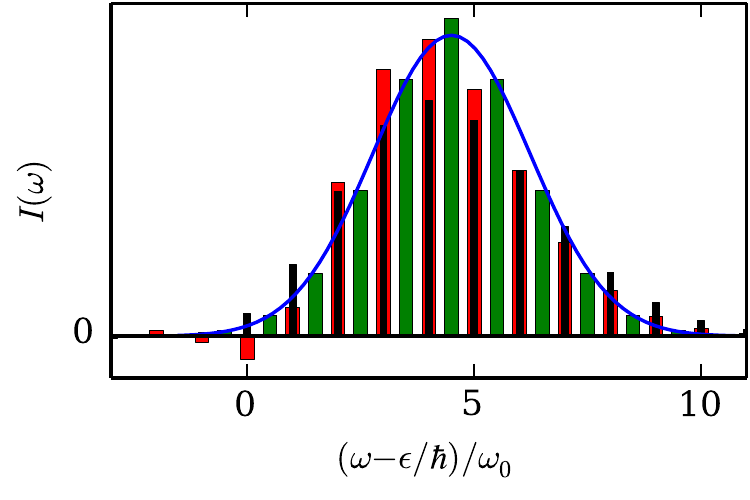}
	\caption{Vibronic spectrum of the one-dimensional harmonic system
	with parameters $2\kappa^2/\hbar m\omega_0^3=9/2$, $\beta\hbar\omega_0=3$ and $\beta\epsilon\gg1$.
	The exact result (black), DCL (green) and ACL (red)
	are a set of delta functions
	represented by the vertical bars of height proportional to the integral over the delta peaks.
	The continuous SCL spectrum is in blue.}
	\label{fig:classical}
\end{figure}

In \fig{classical}, the spectra
are shown for these classical approximations for a system
which describes a typical molecular situation.
It is common at room temperature for $\beta\hbar\omega_0$ to be greater than 1
and thus
the classical approximations
are not expected to 
describe the exact result faithfully.

Nonetheless, the ACL results appear surprisingly accurate and are able to describe the location of the peaks in the discrete spectrum,
which is a purely quantum-mechanical effect.
This is a consequence of the fact that semiclassical dynamics 
is able to predict the correct energy levels of a harmonic oscillator \cite{GutzwillerBook},
but this will not be the case  
for a more general system with anharmonic potentials.
The heights of the peaks are however not correct
for either ACL or DCL,
and the envelopes of all the classical approaches
are obviously not broad enough
due to the incorrect description of the quantum thermal distribution in the ground state.

Unlike the previous classical approaches, the NRPMD method, like standard RPMD \cite{RPMDcorrelation},
provides an approximation to Kubo-transformed correlation functions.
The Kubo-transformed dipole-dipole autocorrelation function,
$\tilde{C}_{\mu\mu}(t)$,
is defined by \eqn{CAB}.
Given this correlation function, 
the spectrum can be computed
using \eqs{spectrum} and (\ref{Kubo}).
We expect that NRPMD will be able to improve upon some of the deficiencies of the classical approaches.
In particular, it will be able to treat
the quantum distribution correctly
and should thus lead to more accurate results.

There is a subtle difference between the NRPMD approach and the classical approaches given above.
In the classical approaches, the initial distribution was 
assumed to have been prepared in the ground state,
such that only the absorption process was described.
However,
in the NRPMD methodology, the initial state is a thermal distribution
over both potential-energy surfaces.
It is not easy to separate the Kubo-transformed correlation function
into terms describing absorption from the ground state
from emission from the excited state,
and thus they are both studied together.
One should therefore be careful not to directly compare the NRPMD results with the classical approximations
unless the energy gap is large enough,
such that the excited state is naturally unpopulated.

For this system, the NRPMD Hamiltonian can be written
\begin{align}
	\label{HJ}
	\mathcal{H}_N %
	= \frac{|\mathbf{p}|^2}{2m} + U_N(\mathbf{x})
	+ \sum_{i=1}^N \sum_{n=0}^1
		\bar{J}_{i n} V_n(\mat{x}_i) \, ,
\end{align}
where 
the canonical variables are $\{\mat{x}_i\}$, $\{\mat{p}_i\}$,
$\{\frac{\hbar}{\Ne} J_{i\alpha n}\}$ and $\{\theta_{i\alpha n}\}$.
The nuclear trajectories are obtained
by solving Hamilton's equations of motion:
\begin{subequations}
\begin{align}
	\dot{\mat{x}}_i &= \frac{\mat{p}_i}{m}
\\
	\dot{\mat{p}}_i &=
	- \pder{U_N}{\mat{x}_i}
	- \sum_{n=0}^1 \bar{J}_{in} \pder{V_n}{\mat{x}_i}
\\
	\dot{J}_{i\alpha n} &= 0
\\
	\dot{\theta}_{i\alpha n} &= \frac{1}{\hbar} V_n(\mat{x}_i)
\end{align}
\end{subequations}
and thus
$\bar{J}_{in}$ is constant,
which makes the propagation of trajectories particularly simple.
Each bead is coupled to its neighbours by ring-polymer springs but follows a different effective potential-energy surface
depending on the values of $\bar{J}_{in}$.

The observable for the transition-dipole moment in the mapping representation is given by
\begin{align}
	\bar{\mu} &= \frac{1}{N\Ne} \sum_{i=1}^N \sum_{\alpha=1}^\Ne \mu_{i\alpha}
	\\
	\mu_{i\alpha} &= X_{i\alpha 0} X_{i\alpha 1} + P_{i\alpha 0} P_{i\alpha 1} \, ,
\end{align}
whose time dependence can be written in action-angle variables as
\begin{align}
	\mu_{i\alpha}(t) &= \sqrt{2J_{i\alpha 0}+1} \sqrt{2J_{i\alpha 1}+1} \cos{\vartheta_{i\alpha}(t)} \, ,
\end{align}
where $\vartheta_{i\alpha}(t) = \vartheta_{i\alpha}(0) + \frac{1}{\hbar} \int_0^t \big[V_1(\mat{x}_i(t)) - V_0(\mat{x}_i(t))\big] \, \rmd t$.

The NRPMD approximation to the Kubo-transformed dipole autocorrelation function is 
\begin{multline}
	\label{Clong}
	\tilde{C}_{\mu\mu}(t)
	\approx
	\iiiint
	\frac{1}{N\Ne}
	\sum_{k=1}^N
	\sum_{\xi=1}^\Ne
	\left(\left[\prod_{i=1}^N \prod_{\alpha=1}^\Ne \Phi_{i\alpha}\right]_{01} + \left[\prod_{i=1}^N \prod_{\alpha=1}^\Ne \Phi_{i\alpha}\right]_{10}\right)
	\mu_{k\xi}(t)
	\\ \times
	\eu{-\sum_{i\alpha n} (2J_{i\alpha n}+1) -\beta_N[|\mathbf{p}|^2/2m + U_N(\mathbf{x})]}
	\, \rmd\mathbf{x} \, \rmd\mathbf{p} \, \rmd\mathbf{J} \, \rmd\mathbf{\vartheta} \, ,
\end{multline}
where the integration range of $J_{i\alpha n}$ is $[-\thalf,\infty)$, and of $\vartheta_{i\alpha}$ is $[-2\pi,2\pi)$.
Note that the integration variables are the initial conditions (at $t=0$) for the trajectories.

In \secref{analysis}, we were able to integrate over the cyclic variables $\Theta_{i\alpha}$.
Here, 
due to the simple uncoupled form of the Hamiltonian, \eqn{HJ},
we can go one step further, as due to the absence of nonadiabatic coupling, 
$\vartheta_{i\alpha}$ is also cyclic
and can be integrated out analytically
to obtain the simpler expression
\begin{align}
	\label{CNRPMD}
	\tilde{C}_{\mu\mu}(t)
	\approx
	\iiint 
	\frac{1}{N\Ne}
	\sum_{k=1}^N
	\sum_{\xi=1}^\Ne
	\rho_{k\xi}(\mathbf{x},\mathbf{p},\mathbf{J})
	\cos\left( \frac{1}{\hbar} \int_0^t \big[V_1(\mat{x}_k(t)) - V_0(\mat{x}_k(t))\big] \rmd t \right)
	\rmd\mathbf{x} \, \rmd\mathbf{p} \, \rmd\mathbf{J} \, ,
\end{align}
which gives the same result as \eqn{Clong} but is more efficient to compute.
The distribution is defined by the terms %
\begin{align}
	\rho_{k\xi}(\mathbf{x},\mathbf{p},\mathbf{J})
	=
	\left( \left[ \prod_{i=1}^N \prod_{\alpha=1}^\Ne \mat{R}_{i\alpha} \right]_{01} + \left[\prod_{i=1}^N \prod_{\alpha=1}^\Ne \mat{R}_{i\alpha} \right]_{10} \right)
	\eu{-\sum_{i\alpha n} (2J_{i\alpha n}+1) -\beta_N[|\mathbf{p}|^2/2m + U_N(\mathbf{x})]} \, ,
\end{align}
where if $i=k$ and $\alpha=\xi$,
\begin{subequations}
\begin{align}
	\mat{R}_{i\alpha} &= (2J_{i\alpha 0}+1) (2J_{i\alpha 1}+1) \big(M_{i0}(2J_{i\alpha 0}+1) + M_{i1}(2J_{i\alpha 1}+1) \big) \begin{pmatrix} 0 & M_{i0} \\ M_{i1} & 0 \end{pmatrix},
\end{align}
and otherwise,
\begin{align}
	\label{Rotherwise}
	\mat{R}_{i\alpha} &= 2 \begin{pmatrix} M_{i0}^2 (2J_{i\alpha 0}+1)^2 & 0 \\ 0 & M_{i1}^2 (2J_{i\alpha 1}+1)^2 \end{pmatrix}.
\end{align}
\end{subequations}

\Eqn{CNRPMD} is the main result of this section
and describes a new approach for simulating vibronic spectra in complex molecular systems.
The function can be computed using standard Monte Carlo techniques,
sampling initial conditions from the integrand at $t=0$ and propagating trajectories according to the equations of motion.
Note that unlike more general applications of NRPMD, the distribution here is positive definite and thus does not suffer from a sign problem at all.
This is because we have integrated explicitly over all angle coordinates, $\theta_{i\alpha n}$, leaving only positive terms.

All of the results in \secref{accuracy} apply to the NRPMD dynamics here
and thus, in the $\Ne\rightarrow\infty$ limit,
the mapping Hamiltonian is unique
and inverted potentials are excluded at $t=0$.
Because there is no nonadiabatic coupling
and all $J_{i\alpha n}$ variables are constant,
no zero-point energy leakage can occur
and thus the inverted potentials are excluded at all times.

This NRPMD method shares some similarities with the classical approaches \eqn{Ccl}.
Although the distributions and dynamics are different,
the observable is in both cases an oscillating function of the instantaneous energy gap between the ground and first-excited states.
However, in contrast with the classical approaches, the NRPMD distribution causes each bead to follow different effective potential surfaces.
Due to the simple form of \eqn{Rotherwise} and the consequences of the
central-limit theorem discussed in \secref{accuracy},
the ring polymer is seen to be initialized with some beads in state $n=0$ and the remainder in state $n=1$.
The hops occur between the last and first bead and also on the $\alpha=\xi$ mapping variable of bead $i=k$.

Although there were two classical-dynamics approaches presented above, DCL and ACL,
one could imagine a whole family of methods with dynamics on the effective potential surface
\mbox{$U(\mat{x}) + (1-\lambda) V_0(\mat{x}) + \lambda V_1(\mat{x})$} where $0\le \lambda \le 1$.
DCL and ACL are represented by $\lambda=0$ and $\lambda=\half$
but there is no derivation which favours any particular value of $\lambda$ over the others.
NRPMD resolves the ambiguity and has only one form, which was derived from the Kubo-transformed correlation function
and can be thought of as including contributions from all values of $\lambda$.

For the harmonic system described above, the equations of motion can be obtained 
and all integrals can in principle be performed analytically.
Although this process leads to a closed-form result, it is rather long and complicated and does not offer much insight.
We will instead study the results obtained by computer algorithms for a specific case.

\begin{figure}
	\centering
	\includegraphics{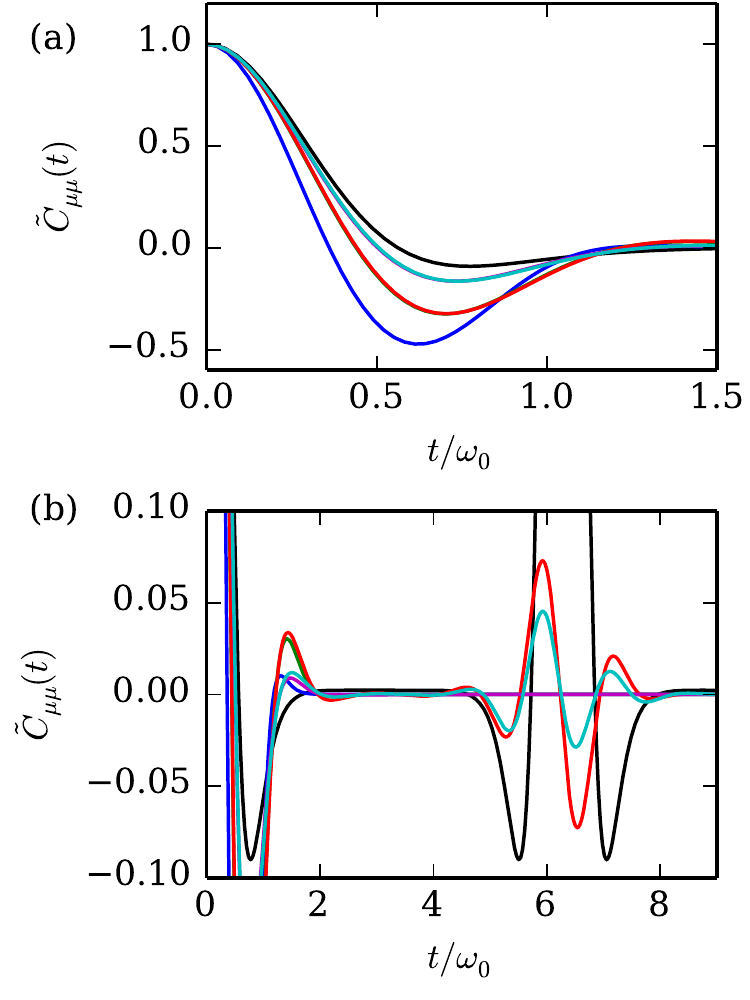}
	\caption{Correlation function for the one-dimensional harmonic system
	with parameters $2\kappa^2/\hbar m\omega_0^3=9/2$, $\beta\hbar\omega_0=3$ and $\epsilon=0$.
	The exact result is shown by a black line
	and the
	NRPMD results are shown
	with
	blue ($N=1$, $\Ne=1$), green ($N=1$, $\Ne=8$), red $(N=1$, $\Ne=512)$,
	magenta ($N=16$, $\Ne=1$) and cyan ($N=16$, $\Ne=128$) lines.
	Each function has been normalized so as to start at 1.
	Note that at short times, the red and cyan lines obscure the green and magenta lines.
	Part (b) shows a zoomed-in representation of same functions for longer times than in part (a).
	Only for the red and cyan lines, which have a large value of $\Ne$,
	does the recurrence appear,
	although it is still much weaker than in the exact case.
	}
	\label{fig:twoCs}
\end{figure}

In \fig{twoCs}, the Kubo-transformed correlation functions are shown for various values of $N$ and $\Ne$\@.
There they are compared with the exact result, which
is periodic with a period of $2\pi/\omega_0$.
Choosing $N=16$ and $\Ne=1$ is enough to converge the NRPMD results for the range of times shown in the upper part of the figure,
i.e. for the first peak only.
It is seen that with the larger values of $N$, the results agree much better with the exact case.
This agreement cannot be obtained with $N=1$, even for large $\Ne$, although some improvement is found over the $\Ne=1$ case.

As shown in the lower part,
even with $N=16$, the recurrence of the correlation function
at $t=2\pi/\omega_0$ is not apparent for small values of $\Ne$.
Interestingly, 
increasing $\Ne$ does seem to improve the situation
and and the peaks appear slowly.
The results presented here are not converged with respect to $\Ne$ and the recurrence peak shows signs of continuing to grow as the
number of mapping variables is increased.
Reaching the converged limit at long times was unfortunately not possible in a reasonable computational time,
although the short-time part was converged relatively easily.

\begin{figure}
	\centering
	\includegraphics{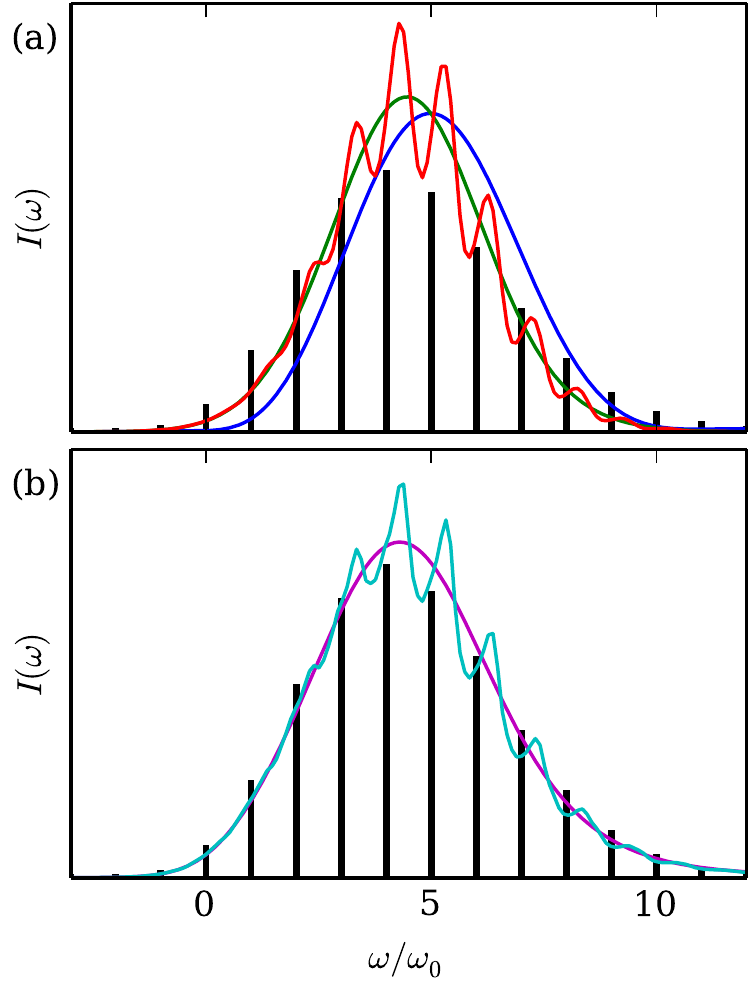}
	\caption{Vibronic spectra obtained from the NRPMD correlation functions
	for the system described in \fig{twoCs}.
	The exact result is a set delta functions represented by the vertical black lines of height equal to the integral over the delta peaks.
	The NRPMD results are shown with the same colour scheme as in \fig{twoCs} but
	separated into two panels for (a) $N=1$ and (b) $N=16$.
	}
	\label{fig:oneD}
\end{figure}

The spectra obtained from these correlation functions are shown in \fig{oneD}.
With small values of $\Ne$, there are no discrete peaks in the spectrum,
but the agreement with the envelope of the exact spectrum
is seen to be much improved for $N=16$
when compared with the classical case of $N=1$.
This is because the classical distribution of $x$ in the ground state
is narrower than the quantum distribution,
which can be seen from a comparison of the standard deviations of the classical Boltzmann distribution, $1/\sqrt{\beta m \omega_0^2}$,
with that of the thermal Wigner distribution,
$\sqrt{\hbar/2m\omega_0\tanh(\beta\hbar\omega_0/2)}$.
This is also the reason that the envelopes of the classical results in \fig{classical} are too narrow.

The spectrum for the $N=\Ne=1$ case is shifted incorrectly to higher frequencies.
This error can be corrected by increasing $\Ne$ to at least 8 and thus
suggests that the cause of the error is related to the problems of the non-unique choice of Hamiltonian and
inverted potentials,
which are avoided by using larger $\Ne$ values as described in \secref{accuracy}.
Using $N=16$ and $\Ne\ge1$ also corrects for this problem as the mapping variables on the different ring-polymer beads are also averaged.

With a large value of $\Ne$,
the spectrum remains continuous but peaks appear whose locations seem to match reasonably well with the exact spectrum.
A similar effect is observed when $N$ becomes very large while keeping $\Ne=1$,
but this is more difficult to compute and is not shown.

Note that the correlation function has to be computed for longer times than presented in \fig{twoCs}
in order to observe this effect
as it is the partial recovery of periodicity which causes it.
The better the recurrences are captured, the longer trajectory will have to be propagated and the more pronounced the peaks will be.
In larger systems, where the density of vibrational states is higher,
the correlation function will be dampened after a short time and
the discrete peaks will be broadened by decoherence.
As we have shown, it is possible to converge the NRPMD correlation function at short times
and it is thus expected to be an practical method which provides a good approximation to the exact spectrum.

We finally note that there are no resonance problems here of the sort which contaminate RPMD simulations of vibrational spectra.
This is because we have used a dipole operator which is independent of position.
If we were to generalize the method to include position-dependent operators \cite{Wang2004absorption}, artificial resonances may occur
and we would require a thermalized ring-polymer approach \cite{Rossi2014resonance} to remove them.

\section{Conclusions}

We have described a generalization of the NRPMD method
which assigns many mapping variables to each ring-polymer bead.
The advantages of this approach are that convergence can be reached more efficiently without 
requiring many computations to the potential-energy surface.
An extreme example would be the choice of $N=1$ with $\Ne>1$.
This would be appropriate for describing the dynamics of a heavy classical nucleus which does not exhibit nuclear delocalization.
The use of many mapping variables should however improve the results with respect to the
classical mapping approach \cite{Meyer1979nonadiabatic,Mueller1998mapping,Stock2005nonadiabatic}.

We have addressed the sign problem to improve the efficiency of an NRPMD simulation
in a general way by integrating analytically over cyclic variables.
Although this problem is reduced, it still exists in certain situations
and limits the number of mapping variables that can be considered
in practical computations.
In certain cases, the sign problem can be completely removed,
such as in the vibronic spectra calculations described in this paper.

Studying the limit of $\Ne\rightarrow\infty$ reveals some interesting behaviour
which suggests that some of the problems of the classical mapping approach are avoided
by NRPMD.
For instance, the total population of the electronic states is automatically forced to be exactly 1
by the appropriate Boltzmann distribution
which removes the artificial dependence on the choice of mapping Hamiltonian.
Also, the initial average populations cannot be negative which
avoids the problem of inverted potentials.

A new application to simulating vibronic spectra is described
which shows promising results for the system studied here.
An exact initial quantum distribution is included in the NRPMD description
and thus leads to a good description for the envelope of the spectrum.
For small systems with discrete spectra,
there is also evidence that the NRPMD approach is able to approximately predict the location and height of the peaks
corresponding to vibronic transitions.
In large complex problems where nuclear decoherence dominates,
the peaks will be washed out and the approximation should be excellent.

\section{Acknowledgements}
This paper is dedicated to Professor Lorenz Cederbaum on the occasion of his 70th birthday.
We thank the undergraduate project students
W.\ David K.\ Jung %
and Christian Hertlein, %
who tested some early ideas which led to the results presented in this paper.
This work was supported by the Alexander von Humboldt Foundation
and a European Union COFUND/Durham Junior Research Fellowship.

\appendix

\section{Appendix}
\label{sec:appendix}

Here we give the closed-form integrals for 
\eqn{Ti}.
To simplify the notation,
each variable refers to a single instance of the mapping variables
and has its index $i\alpha$ missing.

\begin{align*}
	[\mat{T}]_{0000}
		&= \frac{1}{32} (2J_0+1)^2 (8\pi - 4|\vartheta| + \sin|4\vartheta|) \\
	[\mat{T}]_{0001} = [\mat{T}]_{0010}
		&= \frac{1}{32} (2J_0+1)^{3/2}(2J_1+1)^{1/2} \big((8\pi - 4|\vartheta|)\cos\vartheta - 5 \sin|\vartheta| +3 \sin|3\vartheta|\big) \\
	[\mat{T}]_{0100} = [\mat{T}]_{1000}
		&= \frac{1}{32} (2J_0+1)^{3/2} (2J_1+1)^{1/2} \big((8\pi - 4|\vartheta|)\cos\vartheta + 7 \sin|\vartheta| - \sin|3\vartheta|\big) \\
	[\mat{T}]_{0011} = [\mat{T}]_{1100}
		&= \frac{1}{32} (2J_0+1)(2J_1+1) \big((8\pi - 4|\vartheta|)(2 - \cos2\vartheta) + 2\sin|2\vartheta|\big) \\
	[\mat{T}]_{0110} = [\mat{T}]_{0101} = [\mat{T}]_{1001} = [\mat{T}]_{1010} 
		&= \frac{1}{32} (2J_0+1)(2J_1+1) \big((8\pi - 4|\vartheta|)\cos2\vartheta + 2\sin|2\vartheta|\big) \\
	[\mat{T}]_{0111} = [\mat{T}]_{1011}
		&= \frac{1}{32} (2J_0+1)^{1/2}(2J_1+1)^{3/2} \big((8\pi - 4|\vartheta|)\cos\vartheta + 7 \sin|\vartheta| - \sin|3\vartheta|\big) \\
	[\mat{T}]_{1101} = [\mat{T}]_{1110}
		&= \frac{1}{32} (2J_0+1)^{1/2} (2J_1+1)^{3/2} \big((8\pi - 4|\vartheta|)\cos\vartheta - 5 \sin|\vartheta| + 3 \sin|3\vartheta|\big) \\
	[\mat{T}]_{1111}
		&= \frac{1}{32} (2J_1+1)^2 (8\pi - 4|\vartheta| + \sin|4\vartheta|)
\end{align*}

\end{document}